\documentclass[reprint,aps,prl,showpacs]{revtex4-1}
\usepackage{graphicx,epsfig}
\usepackage{epstopdf}
\usepackage{bm}

\newcommand{\3}{$^3$He}
\newcommand{\A}{$^3$He-{A}}

\newcommand{\etal}{{\it et al.}}
\newcommand{\bH}{\mathbf{H}}

\newcommand{\bl}{\hat{\mathbf{l}}}

\newcommand{\Fred}{Fr\'{e}edericksz }
\newcommand{\Omhat}{\mathbf{\hat{\Omega}_0}}

\begin{document}
\title{Chirality of superfluid \3-A}
\author{P.\,M.\,Walmsley and A.\,I.\,Golov}
\address{School of Physics and Astronomy, The University of Manchester, Manchester M13 9PL, UK}
\date{\today}
\begin{abstract}
We have used torsional oscillators, containing disk-shaped slabs of superfluid \A, to probe the chiral orbital textures created by cooling into the superfluid state while continuously rotating. Comparing the observed flow-driven textural transitions with numerical simulations of possible textures shows that an oriented monodomain texture with $\bl$ {\it antiparallel} to the angular velocity $\mathbf{\Omega_0}$ is left behind after stopping rotation. The bias towards a particular chirality, while in the vortex state, is due to the inequivalence of energies of vortices  of opposite circulation. When spun-up from rest, the critical velocity for vortex nucleation 
depends on the sense of rotation, $\hat{\mathbf{\Omega}}$, relative to that of $\bl$. A different type of vorticity, apparently linked to the slab's rim by a domain wall, appears when $\hat{\mathbf{\Omega}} = \bl$. 
\end{abstract}

\maketitle
Chiral superconductors and superfluids (with spontaneously broken time-reversal and parity symmetry) attract interest because of many properties that depend upon the chosen sense of orbital rotation and of the possibility to have topologically-protected quantum states \cite{chiralSC,chiralSC2}. To investigate these, a mono-domain texture with a known orientation of the order parameter is required. So far, even in field-cooled samples of chiral p-wave superconductor Sr$_2$RuO$_4$, domain sizes not larger than $\sim 1$\,$\mu$m are reported \cite{Kidwingira2010}, while the very existence of domain walls separating ground states (vacua) with opposite chiralities was only demonstrated indirectly \cite{indirectDomainWalls}. Many other desirable proofs of the chiral character of Sr$_2$RuO$_4$ (such as the inequivalence of the critical fields and structures of defects for opposite directions of magnetic fields \cite{Tokuyasu1990,Sauls2009}) are yet to be made -- and require finding some means of obtaining monodomain samples in the first place. 

\A\ is a chiral p-wave superfluid \cite{VW} with the unit vector $\bl$ describing the local direction of the coherent orbital momentum of Cooper pairs.   
 When \A\ in zero magnetic field is confined between two parallel walls, normal to $\hat{\mathbf{z}}$ and separated by a distance $D$, $\bl$ is forced normal to the boundaries, leading to a two-fold degenerate ground state with a uniform $\bl$ texture (i.\,e. $\bl=\pm\hat{\mathbf{z}}$). These chiral textures are the time reversed states of each other. 
 This quasi-2-d \A\ is analogous to the case of anisotropic Sr$_2$RuO$_4$ in which the $\bl$-vector is aligned with the c-axis. Yet \A\ differs by the ability of its order parameter to break off the constrain $\hat{\bf{l}} \parallel \pm \hat{z}$ at macroscopic lengthscales $\leq D$.
While similar to chiral superconductors, \A\ is free of many problems that plague them: it is free of complicated rotational anisotropy and crystal defects; there is no long-range magnetic interaction between different regions of vorticity and surface currents; its large-core vortices-skyrmions experience no bulk pinning and only very weak pinning by container walls -- thus
 allowing them to be completely removed after stopping rotation. 

In the absence of any orientational bias to lift the degeneracy, a transition into the superfluid state produces multiple regions of uniform $\bl$-texture of both chiral types separated by domain walls; such polydomain textures display reduced critical velocities (identical for rotation in opposite directions) for the \Fred transition and vortex nucleation \cite{Fred,Walmsley03}, and also the strong intrinsic pinning of vorticity by the network of domain walls \cite{Walmsley04}. 
 In this Letter, we demonstrate asymmetry with respect to the sense of rotation in monodomain samples of \A\ which we succeeded in creating by cooling \3\ whilst rotating. 
The orientation of the obtained ${\bl}$-texture was found to be actually {\it opposite} to the angular velocity of rotation that was producing the bias.

\begin{figure}[t]
\includegraphics[width=6.5cm]{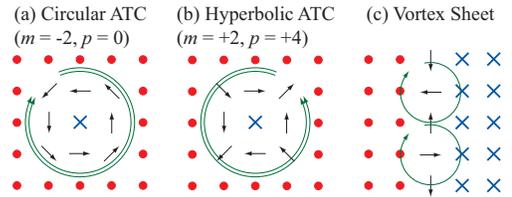}
\caption{(Color online). ${\bl}$-textures of non-singular vortex states in a mono-domain (a, b) and poly-domain (c) quasi-2d $^3$He-A. The circles and crosses represent the chiral uniform textures with $\bl=\hat{\mathbf{z}}$ and $\bl= -\hat{\mathbf{z}}$ respectively. Black arrows show the orientation of the distorted regions where $\bl \perp \hat{\mathbf{z}}$. The quanta of velocity circulation are indicated by the green circular arrows.}
\label{fig1}
\end{figure}

Cooling through the temperature of the second-order phase transition, $T_{\rm c}$, whilst rotating at ${\mathbf{\hat{\Omega}}_0}=\pm\hat{\mathbf{z}}$ seems to be one way to selectively create the chiral vacuum of the lowest free energy \cite{Bevan}. The Zeeman-like interaction, $\propto - {\mathbf{L}_0} \cdot {\mathbf{\Omega}_0}$, due to the tiny intrinsic orbital momentum ${\mathbf{L}_0}$ \cite{VW} would favor $\bl\parallel +{\mathbf{\Omega}}$; however, the effect of quantized vortices should be much greater. As the thermodynamic first critical field is zero in \A, one arrives into the vortex state. 
Superfluid \A\ allows a wide variety of vortex types \cite{Salomaa1987,Parts95} with either singular or continuous cores, and hence several different vortex states. In our conditions (thick slab, zero magnetic field, substantial angular velocities) \cite{Hu1978}, the continuous vortices-skyrmions of Anderson-Toulouse-Chechetkin (ATC) \cite{AT, Chechetkin, Salomaa1987} have the lowest energy. These are metastable $\bl$-textures with the diameter of the soft core $\sim D$, within which the ${\bl}$-vector flips. The circulation of the superfluid velocity round any contour is related to the inhomogeneous ${\bl}$-texture within the contour through the Ho theorem \cite{Ho1978}.
  Panels (a) and (b) in Fig.\,\ref{fig1} show ATC vortices of opposite senses of circulation embedded in the $\bl=\hat{\mathbf{z}}$ ground state. They have $m=-2w$ quanta of circulation, $\kappa \equiv h/2m_3$, where the integer $w$ denotes the winding number of the $xy$ component of $\bl$ inside the core region (taking counterclockwise rotation as positive). The axisymmetric `circular' vortex (a) with $w=+1$ ($m=-2$) posesses an orbital momentum {\it antiparallel} with $\bl=\hat{\mathbf{z}}$, whereas the `hyperbolic' vortex (b) with $w=-1$ ($m=+2$) has its orbital momentum {\it parallel} with $\bl=\hat{\mathbf{z}}$. One can think of their soft cores as of circular domains of the time-reversed state ($\bl=-\hat{\mathbf{z}}$) surrounded by a domain wall, which is decorated by compact vortex kinks-merons, each with one quantum of circulation. The total circulation is hence $m=-2+p$, where the even integer $p$ gives the number of circulation quanta contributed by kinks: $p=0$ for circular (a) and $p=+4$ for hyperbolic (b) vortices. An extended domain wall between the degenerate time-reversed ground states, decorated with vortex kinks is generally known as the vortex sheet \cite{Parts95} (Fig.\,\ref{fig1}(c)).

The free energy in a frame rotating together with the container and viscous normal component is $F=F_{0}+\pi R^2 nE_{\rm v}-\mathbf{L} \cdot \mathbf{\Omega}$, where  $F_{0}$ is the free energy in the laboratory frame associated with the underlying global $\bl$ texture, $n$ is the areal density of vortices, $E_{\rm v}$ is the energy of a vortex, $R$ is the radius of the container and $\mathbf{L}$ is the total angular momentum. The inequivalence of the two types of ATC vortex suggests that they are capable of providing an orientational bias such that a particular chirality of $\bl$ is preferred, dependent on which type has the lowest energy $E_{\rm v}$. 

We used the approach of Karim\"aki and Thuneberg \cite{Karimaki99} to calculate the free energy for the 2-d ATC vortices. We find that the ratio of energy between the two vortex types is $E_{\rm v}(m=-2)/E_{\rm v}(m=+2)=0.89$, thus predicting that the oriented ground state texture with $\bl= -\mathbf{\hat{\Omega}}_0$ should be favored. Basically, these calculations confirm the expectation that the smoother ${\bl}$-texture of the soft core of the circular vortex ($p=0$) results in a lower vortex energy than the kinked core of the hyperbolic vortex ($p=4$). As kinks generally add energy, the vortex with fewer kinks (smaller $|p|$) should have the lowest energy. The formula $m=-2+p$ then suggests the equilibrium vortex state is that with $p=0$, i.\,e. $m=-2$.  

We investigated textures, using the torsional oscillator (TO) technique, in \3-A inside two different disk-shaped cavities, both with radius $R=5.0$\,mm, but with thicknesses $D \approx 0.26$ and 0.41\,mm. In this geometry, there are two additional factors that affect the texture: the edges of the disk and the filling line (of radius 0.4\,mm) that arrives via the torsion stem into the disk on its axis. While the texture inside the fill line has negligible effect on the TO properties, the way it merges with the texture in the slab is important because (as was indeed the case) it might possess an additional two-quantum vortex. The TO resonance frequency and width are sensitive to counterflow velocity, particularly near the rim of the disk, through the changes in the anisotropic density and viscosity of the normal component resulting from flow-induced azimuthal tilting of $\bl$.

The TOs were mounted on a rotating nuclear demagnetization cryostat with the disk's axes aligned with the rotation axis of the cryostat ($\hat{z}\parallel\Omega$). All experiments were at 29.3\,bar pressure and with temperatures in the range 2--2.5\,mK. The motion of the TOs was driven and detected capacitively at a frequency close to the resonant frequencies of $\nu_{\rm R}$= 627 and 674\,Hz and the full width at half maximum (bandwidth) was $\nu_{\rm B}$=0.12\,Hz and 0.25\,Hz for the thin and thick slabs respectively. The viscous penetration depth was $\sim D$; hence, both $\nu_{\rm R}$ and $\nu_{\rm B}$ showed a similar response so we used whichever had the better signal to noise ratio.

\begin{figure}[t]
\includegraphics[width=7cm]{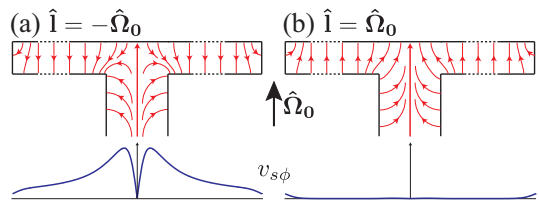}
\caption{(Color online). The two possible textures 
 (mainly the distorted textures at the rim and at the fill line in the centre of our disk-shaped cavities are shown, while the extended uniform textures, $\bl = \pm \hat{\mathbf{z}}$, in between are skipped for clarity) and their associated azimuthal component of superflow, $v_{s\phi}$ (from Eq.\,\ref{eqnvs}). $\bl=-\Omhat$ (left, a) and $\bl=\Omhat$ (right, b).}
\label{fig2}
\end{figure}

To create oriented textures, we cooled through $T_{\rm c}=2.5$\,mK at a rate of $\sim 1$\,$\mu$K\,min$^{-1}$ whilst rotating the cryostat at $\mathbf{\Omega_0}=\pm 0.42\hat{\mathbf{z}}$\,rad\,s$^{-1}$. The magnitude of $\mathbf{\Omega_0}$ was chosen so that the texture in the fill line was the Mermin-Ho type with $1\kappa\Omhat$ of circulation and had $\bl=\Omhat$ on axis. It also ensured that the distance between vortices in the slab, $\simeq\left(\kappa/\Omega_0\right)^{1/2}=0.4$\,mm, is sufficiently large that their soft cores of radius $\simeq D/2$ are well-separated. We thus can treat the texture in the slab as an array of ATC vortices embedded in a  uniform oriented texture. After cooling well below $T_{\rm c}$, the rotation was stopped, and the majority of vortices left the slab. The cryostat was then rotated gently in the opposite direction with $\mathbf{\Omega}=-0.01\Omhat$ rad/s to remove any (typically 3--8) weakly pinned vortices \cite{Walmsley03}.

The two types of oriented monodomain texture $\bl=\pm\Omhat$, possible in our geometry, are shown in Fig.\,\ref{fig2}. If $\bl=\Omhat$ then the texture is the Mermin-Ho type and there will be $+1\kappa\Omhat$ of circulation due to the bending of $\bl$ within a distance $\simeq D$ from the outer perimeter of the slab (Fig.\,\ref{fig2}(b)). Alternatively, if $\bl=-\Omhat$ then the orientation of the texture in the slab is opposite to that in the centre of the fill line, and ``stitching" these textures together will result in a vortex with $+2 \kappa\Omhat$ of circulation trapped in the centre. There will also be $-1 \kappa\Omhat$ contribution that again arises from the bending of $\bl$ at the perimeter (Fig.\,\ref{fig2}(a)). In both of these cases, the azimuthal component of superflow is given by
\begin{equation} \label{eqnvs}
v_{s\phi}=\frac{(1-l_z)\kappa}{2\pi r}.
\end{equation}

\begin{figure}
\includegraphics[width=8cm]{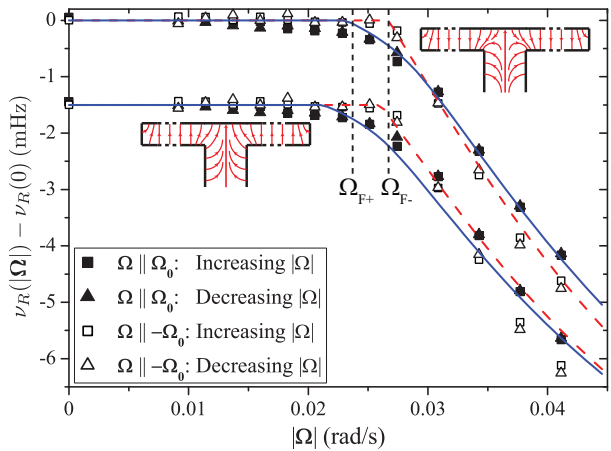}
\caption{(Color online). The shift in $\nu_{\rm R}(\Omega)$ for the $D=0.26$\,mm slab due to the flow-induced \Fred transition for both directions of rotation. The lines (solid blue: $\mathbf{\Omega}\parallel -\bl$, dashed red: $\mathbf{\Omega}\parallel\bl$) are calculated using Eqn.\,\ref{freqshift} for the two possible types of oriented textures (top: $\bl=-\Omhat$ as on Fig.\,\ref{fig2}(a); bottom (shifted by -1.5\,mHz for clarity): $\bl=\Omhat$ as on Fig.\,\ref{fig2}(b)). They are compared with the typical experimental data, $\mathbf{\Omega}\parallel\pm\mathbf{\Omega_0}$ (same data points are also repeated shifted by -1.5\,mHz). $T=0.9 T_{\rm c}$ and $C=-18$\,mHz.}
\label{fig3}
\end{figure}

To figure out which of these textures was formed, 
 we rotated the cryostat at $\Omega>\Omega_{\rm F}\simeq v_{\rm F}/ R$ to induce the flow-driven \Fred transition \cite{Fred} while tracking changes in $\nu_{\rm R}$ and $\nu_{\rm B}$. This was done for both directions of rotation ($\mathbf{\Omega}\parallel\pm\mathbf{\Omega_0}$) but with care not to nucleate vortices. The observed shifts, $\Delta\nu_{\rm R}(\mathbf{\Omega})$ (shown in Fig.\,\ref{fig3}) were reversible with increasing/decreasing $\Omega$. There are reproducible differences for opposite senses of rotation that have been observed in both slabs investigated: for ($\mathbf{\Omega}\parallel\mathbf{\Omega_0}$), the onset of the \Fred transition occurs at a lower $\Omega_{\rm F+}$, and then the TO frequency $\nu_{\rm R}(|\Omega|)$ drops with a shallower slope compared to rotation in the opposite direction (where the transition is noticeably sharper and at a larger $\Omega_{\rm F-}$) \cite{sign} -- so that two $\nu_{\rm R}(|\Omega|)$ quickly change hands before becoming nearly parallel. 

To elucidate the $\bl$ texture obtained in the experiment, we have numerically calculated $\bl(r,z,\mathbf{\Omega})$ in the slab by minimizing the free energy \cite{supp_mat}. The calculated textures were converted into a frequency shift using
\begin{equation} \label{freqshift}
\nu_R(\mathbf{\Omega})-\nu_R(0)=\frac{2C}{\pi D R^4}\int r^2 l_\phi (r,z,\mathbf{\Omega}) \,\mathrm{d}V,
\end{equation}
where $C\propto\rho_s(T)$ is a fitting parameter to provide the correct frequency scaling. In addition, the effective $D$ was increased from 0.26 to 0.28\,mm so that the calculated values of $\Omega_{\rm F}$ coincided with the measurements. 
The calculated frequency shifts are compared to the experimental data in Fig. \ref{fig3}. Clearly the oriented texture with $\bl=-\Omhat$ best reproduces all observed differences in the response of the torsional oscillator for the different directions of rotation. The observed asymmetry results from the combined effect of a trapped double-quantum vortex and the coupling between the applied flow and the non-uniform texture at the perimeter of the slab \cite{supp_mat}. 
We thus confirm our theoretical calculations that $\bl= -{\bf \hat{\Omega}}_0$ is formed by our technique.

\begin{figure}
\includegraphics[width=8cm]{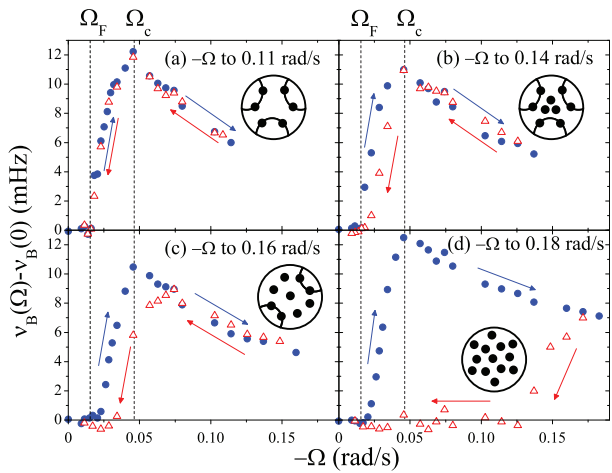}
\caption{(Color online). $\Delta\nu_B$ in the 0.41 mm thick slab for a series of rotation sweeps with $\mathbf{\Omega}\parallel -\mathbf{\Omega_0}$. Solid (open) symbols are for increasing (decreasing) $|\Omega|$. The cartoons indicate how some vorticity is bound to the rim of the container.}
\label{fig4}
\end{figure}

Further evidence for the presence of an oriented texture can be seen in the difference between the critical angular velocities for vortex nucleation, $\Omega_c$, for $\mathbf{\Omega} \parallel\pm\mathbf{\Omega_0}$, and in the case of the thicker slab the subsequent motion of vorticity is entirely different for rotation with $\mathbf{\Omega}\parallel -\mathbf{\Omega_0}$. When an oriented texture is rotated with ${\bf \Omega} ||{\bf \Omega}_0$, circular ATC vortices will be nucleated near the outer rim and then move towards the centre of the disk, forming a vortex cluster, such that $v(R)=v_{\rm c}=\Omega_{\rm c} R$. Upon slowing down, the vortex cluster expands and vortices are able to leave, so a uniform equilibrium distribution of vortices $n=\Omega/\kappa$ is maintained. These rotations with $\mathbf{\Omega}\parallel\mathbf{\Omega_0}$ (which introduced up to $\simeq 500$ ATC vortices) always left behind the original defect-free oriented texture after stopping rotation. The ``$\Omega$-loops'' (like magnetization loops for type-II superconductors) thus show hysteresis in the measured TO properties associated with vortex nucleation at finite $v_{\rm c}$ \cite{Walmsley03} (see Fig.\,\ref{fig4}(d)). For $D=0.26$\,mm (0.41\,mm) we found $v_{\rm c+} \approx 0.6$\,mm\,s$^{-1}$ (0.3\,mm\,s$^{-1}$). We thus interpret $v_{\rm c}(D)$ that decreases with increasing $D$ as $v_{\rm c} \approx 4 v_{\rm F}\propto D^{-1}$  
as the critical velocity for vortex nucleation near the disks edge corners where the $\bl$-texture is severely distorted due to the boundary conditions at the walls \cite{erratum}.  
These critical velocities are consistent with vortices having {\it continuous} cores, for which $v_{\rm c} \sim \kappa/D \sim 0.2$\,mm\,s$^{-1}$ \cite{Walmsley03}, and not with those having {\it singular} cores with $v_{\rm c} \sim \kappa/ \xi_{\rm d} \sim 70$\,mm\,s$^{-1}$ \cite{Ruutu97} ($\xi_{\rm d} \approx 10$\,$\mu$m is the dipole length above which the spin and orbital degrees of freedom are coupled). 

Fast rotation with $\mathbf{\Omega}\parallel-\mathbf{\Omega_0}$ is expected to nucleate hyperbolic ($m=+2,p=+4$) ATC vortices, which when far from the rim will behave the same way as the previously discussed circular ATC vortices. This appears to be the case for the 0.26\,mm slab, where we observe similar $\Omega$-loops albeit with reduced $v_{\rm c-}\simeq 0.4$ mm/s. The oriented texture still remained after rotation is stopped. However, the behavior for the 0.41\,mm slab is very different: even though there are clear evidences of vortex nucleation above $v_{\rm c-}\simeq 0.25$\,mm/s ($\Omega_{\rm c-}\simeq 0.05$\,rad/s), the $\Omega$-loops are still non-hysteretic! Only above $\Omega \simeq 0.14$\,rad/s does the hysteresis gradually set in (indicating that the nucleated vortices stay in the central cluster). Fig.\,\ref{fig4} shows changes in $\nu_B$ for a series of rotation sweeps to progressively higher $\Omega$. $\nu_{\rm B}$ increases when $\Omega>\Omega_{\rm F-}$ but then begins to decrease when $\Omega_{\rm c-}$ ($v_{\rm c-}\simeq0.25$\,mm\,s$^{-1}$) is reached due to the creation of vorticity ($v_{\rm c-}\propto D^{-1}$ for this direction of rotation as well). If the rotation was slowed after reaching a maximum of 0.11\,rad/s as shown in Fig.\,\ref{fig4}(a) (corresponding to introducing $\simeq140\kappa$) then there is no hysteresis at all. If the maximum angular velocity is now increased to 0.14\,rad/s (Fig.\,\ref{fig4}(b)) then there is some slight hysteresis during the final part of the deceleration. The next rotation (Fig.\,\ref{fig4}(c)) has a slightly higher $\Omega_{\rm F-}$ consistent with $\simeq 20\kappa$ being trapped from the previous rotation sweep. Unlike vortices introduced by rotations in the original direction, ${\bf \Omega} || {\bf \Omega}_0$, this vorticity is now strongly pinned \cite{Walmsley04}, and cannot be removed by gentle rotation in the opposite direction. It also seems that defects had been irreversibly introduced into the texture as the behavior for small $\Omega$ $(\Omega_{\rm F-} < \Omega < \Omega_{\rm c-})$ was changed and no longer corresponded to the ones shown in Fig.\,\ref{fig3}. Further rotation to higher angular velocities introduces more hysteresis (Fig.\,\ref{fig4}(c\&d)). The fact that no hysteresis is seen just above $\Omega_{\rm c-}$ means that vortices can be reversibly removed upon (even small) deceleration, which is extraordinary!  

We suggest a speculative scenario  that the non-hysteretic vortex behavior results from vorticity that is bound to the disk's rim. This could occur when a crescent-shaped domain of $\bl=\Omhat$ is formed. Its domain wall will have a reduced critical velocity for the nucleation of vortex kinks \cite{Kopu00} and can either become decorated with four vortex kinks and detach from the wall forming an individual $m=+2,p=+4$ ATC vortex which will migrate towards the centre of the disk (as apparently happens in the thinner slab), or it can stretch and become further decorated with vortex kinks forming a vortex sheet (Fig.\,\ref{fig1}(c)). The ends of the vortex sheet will remain attached to the rim while the vortex kinks will be attracted towards the centre but this will be resisted by the tension of the domain wall. The non-hysteretic behavior probably results from the nucleation of multiple vortex sheets, such a vortex configuration shows very little hysteresis in bulk \A\ \cite{Eltsov02}. The sheets are able to shrink and disappear upon deceleration, until a second critical velocity ($\simeq 0.14$ rad/s) is reached when certain domain walls and kinks remain after rotation is stopped, perhaps due to pinning. Further rotations then show more hysteresis as the kinks are no longer bound to the rim and move towards the centre of the disk. The fact that, following the rotation, at $\Omega$ well-above $\Omega_{\rm c-}$, in the direction opposite to that of ${\bf \Omega}_0$  the initially monodomain texture is permanently ruined (as judged by the \Fred transition) also tells that domain walls have been introduced along with vortex nucleation. This seems to be a natural process of gradual replacement of the metastable vortex state with hyperbolic vortices in $\bl || {\bf \Omega}$ texture by that with circular vortices in $\bl ||- {\bf \Omega}$ texture.

By breaking the time-reversal symmetry of the two competing ground states of chiral \A\  by rotation at ${\bf \Omega}_0$, we have created large monodomain samples of \A\ in a slab geometry and, for the first time, determined their orientation $\bl$ as a function of the bias: $\bl || -{\bf \Omega}_0$. This orientation is opposite to that expected from the interaction with the intrinsic orbital moment, but is due to the differences in the structures of vortex-skyrmions with opposite senses of circulation. It is hence the $-2$ in the formula $m=-2+p$ that explains that the lower-energy vortices with circular core ($p=0$) are embedded into the ${\bl}$-texture oriented {\it opposite} to the sense of initial rotation.
In chiral superconductors, thanks to the inequivalence of energies of vortices with opposite senses of circulation \cite{Sauls2009}, it might also be possible to create monodomain ${\bl}$-textures by a similar technique (i.\,e. slow field-cooling into the vortex state in a substantial magnetic field up to $H\sim H_{\rm c2}$); this will result in $\bl || - \bH$ orientation which is opposite to the one favored by field-cooling experiments with small fields $H \ll H_{\rm c1}$ \cite{Kidwingira2010}.

We acknowledge discussions with H.\,E.~Hall and
the contribution of S.~May in the construction of the
experiment. Support provided by EPSRC under GR/N35113, EP/E001009 and through the award of a Career Acceleration Fellowship to PMW (EP/I003738).
\newpage
\section{Supplementary material}

In the accompanying Letter, we describe how torsional oscillators (TO), containing disk-shaped slabs of \A, respond differently when the direction of rotation is reversed  due to the presence of a chiral monodomain $\bl$ texture. In this supplementary note, we calculate $\bl(\bm{\Omega})$ for two different initial textures (shown in Fig. 2 in the Letter and also reproduced in Fig. 1 in this note). The changes in the moment of inertia are calculated and we find that the distinctive response observed in the experiment can be explained with an initial monodomain $\bl$ texture containing a trapped double-quantum ATC vortex.

Superfluid \3 benefits from a detailed microscopic theory \cite{vw}, facilitating calculations that can be compared to experimental observations. We use the London limit, where the order parameter is always that for bulk \A but the container boundaries along with competing orientational effects determine the direction and spatial variation (texture) of the orbital vector $\bl$. We have calculated axisymmetric $\bl$-textures in a slab of \A in the dipole-locked limit. The upper and lower surfaces of the slab are at $z=\pm D/2$ with the outer perimeter at $r=R$.  The $\bl$ texture, $\bl(r,z)$, was parameterized using Euler angles ($\alpha$,$\beta$,$\gamma$) as
\begin{equation}
\bm{\hat{l}}=\sin\alpha\sin\beta\:\bm{\hat{r}}+\cos\beta\:\bm{\hat{\phi}}+\cos\alpha\sin\beta\:\bm{\hat{z}},
\end{equation}
with the third angle, $\gamma$, representing an additional rotation about $\bl$. The boundary condition that $\bl$ must be perpendicular to the slab surfaces for a monodomain texture with $\bl\parallel\bm{\hat{z}}$ was obtained by setting $\alpha=0,\: \beta=\pi/2$ at $z=\pm D/2$ and $\alpha=\pi/2,\: \beta=\pi/2$ at $r=R$. The components of the superfluid velocity, $v_s$, are given by
\begin{eqnarray}
v_{sr}=-\frac{\partial\gamma}{\partial r}-\cos\beta\:\frac{\partial\alpha}{\partial r}\\
v_{s\phi}=\frac{1}{r}(1-l_z+N)\\
v_{sz}=-\frac{\partial\gamma}{\partial z}-\cos\beta\:\frac{\partial\alpha}{\partial z},
\end{eqnarray}
where $N$ is an integer representing the number of circulation quanta that is trapped on the central axis of the slab. We use natural units where $\hbar/2m_3 =1$ throughout this note. The normal component is locked to the container and undergoes solid body rotation, $\bm{v_n}=\Omega r\, \bm{\hat{\phi}}$ where $\Omega$ is the angular velocity of the applied rotation.

The free energy density for dipole-locked textures in the rotating frame \cite{Fetter83} is given by
\begin{widetext}
\begin{equation} \label{fenergy}
f=\frac{1}{2}\rho_s v^2 - \frac{1}{2}\rho_0 (\bm{\hat{l}}\cdot\bm{v})^2 + C\bm{v}\cdot(\bm{\nabla}\times\bm{\hat{l}})-C_0 (\bm{\hat{l}}\cdot\bm{v})\bm{\hat{l}}\cdot(\bm{\nabla}\times\bm{\hat{l}})+\frac{1}{2}K_s^{'}(\bm{\nabla}\cdot\bm{\hat{l}})^2+\frac{1}{2}K_t^{'}[\bm{\hat{l}}\cdot(\bm{\nabla}\times\bm{\hat{l}})]^2+\frac{1}{2}K_b^{'}|\bm{\hat{l}}\times(\bm{\nabla}\times\bm{\hat{l}}|^2 ,
\end{equation}
\end{widetext}
where $\bm{v}=\bm{v_s}-\bm{v_n}$ is the counterflow and we have neglected surface terms. The texture with the lowest free energy obtained under particular conditions will be determined by the competing terms in Eq. \ref{fenergy} along with the boundary condition. The second term favours $\bl\parallel\pm\bm{v}$ whereas the final three terms favor spatially uniform textures. The third and fourth terms are due to the interaction between flow and regions of inhomogeneous texture. This coupling plays a vital role in our case because the TO is primarily sensitive to the non-uniform textures at the outer perimeter and the applied flow due to rotation is also the highest in this region. At $T\rightarrow T_c$ the coefficents of the various terms in Eq. \ref{fenergy} are related by
\begin{equation}
\rho_0=\frac{1}{2}\rho_s=2 C=C_0=\frac{2}{5} K_s^{'}=\frac{2}{5} K_b^{'}=\frac{2}{5} K_t^{'}.
\end{equation}
Using the coefficients \cite{Fetter79} calculated for the temperature at which the measurement was performed produced a marginally better fit to the data, but does not change any of the main features of the calculated TO response. Thus, for simplicity, all of the calculations described below use $T=T_c$. For the infinite slab geometry, the ground state is a uniform texture ($\bl=\bm{\hat{z}}$) when $\bm{v}=0$. The texture in the middle of the slab ($z=0$) will begin to align with the flow when $v\geq v_{F0}=\sqrt{5}\pi / 2D$, known as a \Fred transition. We thus use $\Omega_{F0} = v_{F0}/R$ as a natural unit for angular velocity in the simulations described below. The \Fred transition in a finite disk-shaped slab will occur at $\Omega_{F0}>1$ due to the transition (i.e. the inital tipping of $\bl$ in the azimuthal direction) occurring at a distance $\simeq D$ from the outer perimeter.

In order to determine the equilbrium texture for particular values of $\bm{\Omega}$, and $N$ the free energy was minimized by solving the Euler Lagrange equations,
\begin{equation}
\nabla\cdot\frac{\partial f}{\partial\nabla x}-\frac{\partial f}{\partial x} = 0,
\end{equation}
where $x=\alpha,\beta,\gamma$. The terms in the three Euler-Lagrange equations were found using \textit{Mathematica} and then solved numerically for different values of $\bm{\Omega}$ using \textit{FlexPDE}, a commercial finite-element method solver. The simulations were typically started for $\Omega\simeq2\Omega_{F0}$. We found that using a simple initial guess of $\alpha=0$, $\beta=\frac{\pi}{4}(2-\frac{r}{R})$, $\gamma=0$ rapidly converged to a stable solution. These solutions were then used as an initial guess for subsequent calculations with a slightly different value of $\Omega$. In order to compare the simulations to the experiments, we calculate
\begin{equation}
\Delta I = \frac{2}{\pi D R^4}\int r^2 l_\phi (r,z,\bm{\Omega}) \,\mathrm{d}V,
\end{equation}
which reflects the changes in the moment of inertia due to the anisotropy of the superfluid density following the flow induced re-alignment of $\bl$ ($\Delta I=0$ when $l_\phi =0$ and $\Delta I=1$ when $l_\phi =1$ everywhere). Thus, the small shifts in the resonant frequency and bandwidth of our TO are proportional to changes in $\Delta I$.

There are three effects that can produce different textures when the direction of rotation is reversed. We discuss each in turn below.

Firstly, a monodomain texture in a disk-shaped slab is essentially a Mermin-Ho vortex with a single quantum of circulation. There is thus superflow, $v_s = \frac{1}{r}$, that flows within $\sim D$ from the outer perimeter of the slab where the texture bends radially outwards. This means that the applied counterflow is slightly different ($v=\frac{1}{r}\pm\Omega r$) depending on the sense of rotation and so will produce a splitting of the \Fred transition for the two directions of rotation: $\Delta\Omega_F/\Omega_{F0}\simeq 4D/\sqrt{5}\pi R = 0.03$ for $R/D$=19.23. This is much smaller than the experimentally observed splitting of $\Delta\Omega_F/\Omega_F=0.15$ (Fig. 3 in the main Letter).

Secondly, a similar effect can be produced by trapped vorticity in the centre of the slab, which has superflow, $v_s=\frac{N}{r}$, but unlike the superflow due to the edge texture, this flow affects the whole slab rather than a limited region at the outer perimeter. This gives a splitting $\Delta\Omega_F/\Omega_{F0}\simeq 4ND/\sqrt{5}\pi R$ and so choosing an appropriate value of $N$ will reproduce the observed splitting, however, it cannot explain the very different slopes of the observed frequency shifts for $\Omega>\Omega_{F}$.

Thirdly, there is coupling between regions of non-uniform $\bl$ (where $\bm{\nabla}\times\bl\neq 0$) and flow due to the $C$ and $C_0$ terms in Eq. \ref{fenergy}. For the inital monodomain texture at low velocities $\Omega<\Omega_F$, $\bm{v}\cdot\bm{\nabla}\times\bl$ is non-zero near the outer perimeter but also changes sign depending on the direction of the flow which will also result in splitting of the \Fred transition but can also produce different slopes of the observed frequency shifts for the different directions of rotation as it affects the distance from the outer perimeter where the texture first begins to tip in the azimuthal direction.

\begin{figure*}
\includegraphics[width=17cm]{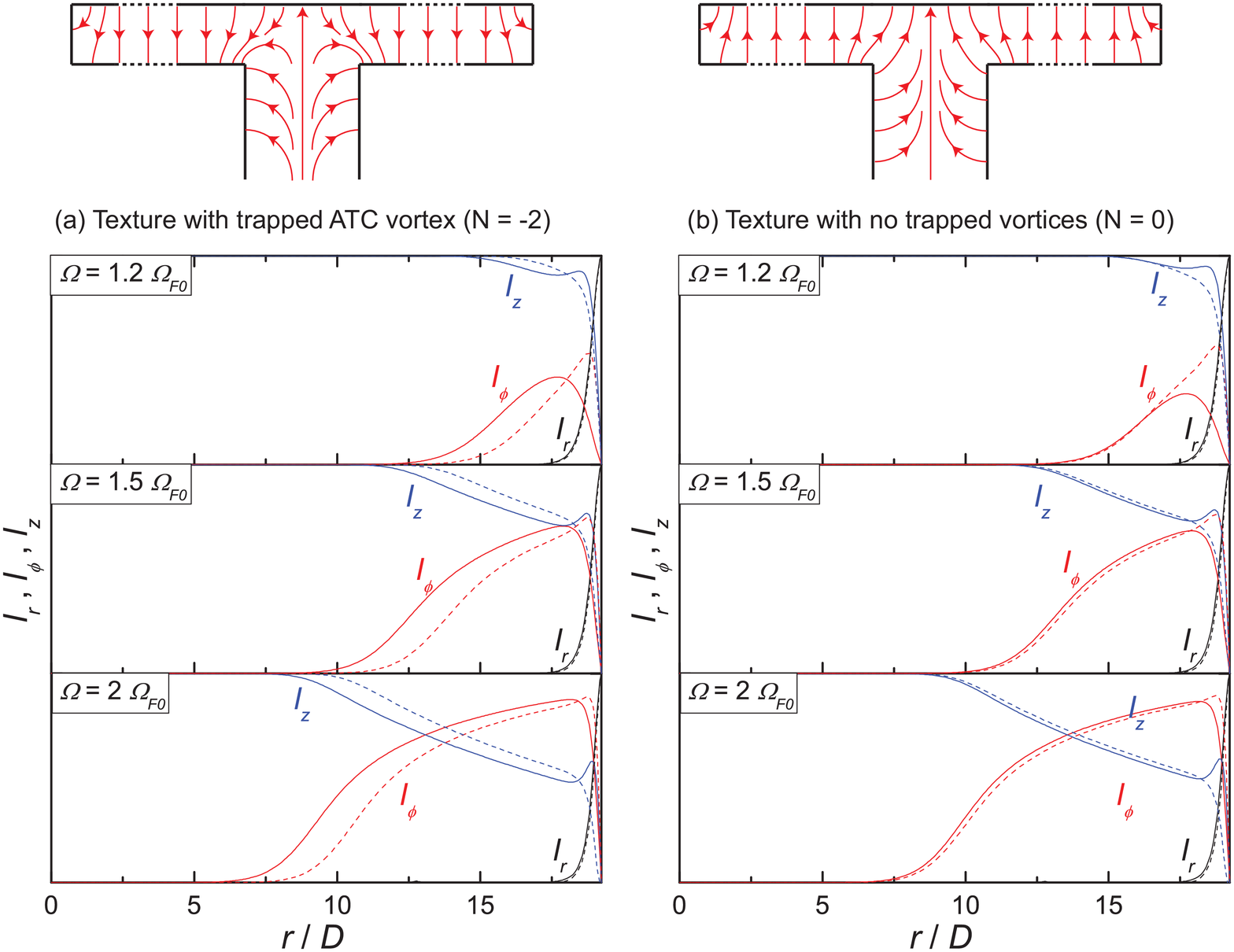}
\caption{(Color online). Components of $\bl$ through the middle ($z=0$) of a rotating slab, with $R/D=19.23$, for three different magnitudes of rotation ($\Omega/\Omega_{F0}=1.2,1.5,2$). The solid and dashed lines are for when $\bm{\Omega}$ is initially parallel and antiparallel to $\bl$ in the monodomain texture respectively. (a) shows the texture that is described in the Letter that best fits our experiment and (b) shows the texture in the absence of the trapped vortex. We take $l_z$ to be postive in both cases to facilitate comparison. The corresponding values of $\Delta I$ for these textures are shown in Fig. \ref{fig2} below.}
\label{fig1s}
\end{figure*}

\begin{figure*}
\includegraphics[width=17cm]{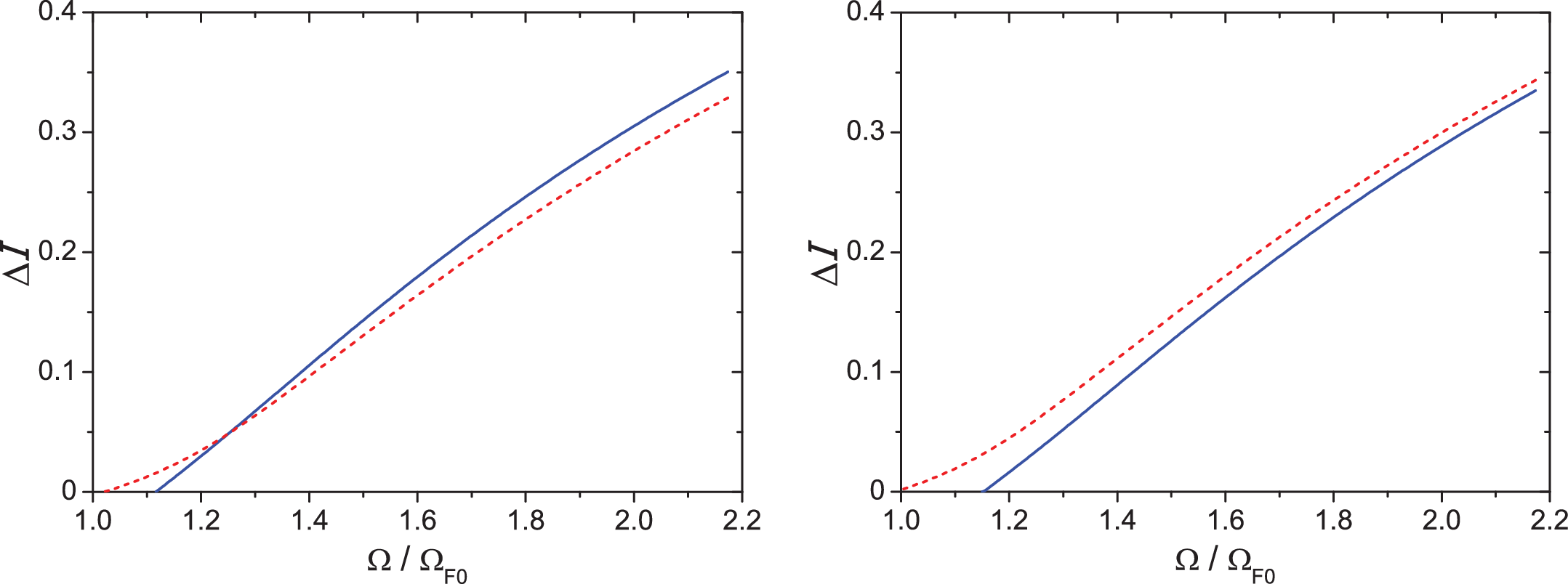}
\caption{(Color online) $\Delta I(\bm{\Omega})$ calculated for the textures shown directly above in Fig. \ref{fig1}. The solid and dashed lines are for $\bm{\Omega}\parallel\bl$ and $\bm{\Omega}\parallel-\bl$ respectively.}
\label{fig2s}
\end{figure*}

The components of $\bl$ across the middle of the slab ($z=0$) going radially outwards from the centre are shown in Fig. \ref{fig1s} for two different initial textures at three values of $|\Omega|$ for both directions of rotation. In general, $\bl$ will begin to tip azimuthally when the local counterflow exceeds $v_{F0}$. This will occur somewhere near the outer perimeter initally and then the belt of tipped texture begins to spread radially inwards as the flow is increased. This is observed for all the configurations of parameters we have tried but the details of the texture differ due to the effects described above. Fig. \ref{fig1s}(a) corresponds to the texture described in the Letter, where a double-quantum ATC vortex is trapped in the centre of the slab due to  `stitching' together the antiparallel textures in the slab and axial fill line. We do not calculate the texture associated with the vortex but include the superflow far from the axis by setting $N=-2$. Fig. \ref{fig1s}(b) shows how the texture is changed when the trapped vortex is removed. We have also tried removing the spontaneous superflow associated with the monodomain texture, but the difference this makes is hardly discernable. We have also calculated the textures with $C=C_0=0$ and found that this produces textures near the outer perimeter that are more symmetric when the direction of rotation is changed. We find that the $C$ and $C_0$ terms are responsible for producing the very different textures near the edge for the different directions of rotation and this leads to different gradients of $\Delta I$ with increasing $\Omega$, particularly near the \Fred transition. The main effect of trapped vorticity is simply to shift the curves relative to each other.  The corresponding $\Delta I(\bm{\Omega})$ are shown in Fig. \ref{fig2s}. Clearly, $\Delta I(\Omega)$ for the texture with the trapped vortex look very similar to the observed frequency shifts. Our conclusion is that only the combination of the trapped vortex and the different coupling between the non-uniform edge texture for the opposing directions of rotation can explain the distinctive features observed in our experiments.

\begin{thebibliography}{99}
\bibitem{Fred} A counterflow ($\mathbf{v}\equiv\mathbf{v_n}-\mathbf{v_s}$) applied parallel to the boundaries favors $\bl$ to be aligned with $\mathbf{v}$. The uniform texture persists until the critical velocity $v_{\rm F}\simeq\sqrt{5}\kappa/4 D$ is reached when $\bl$ begins to tip in the middle ($z=0$) in order to align with the flow \cite{Fetter76}. This flow-driven \Fred transitions has been observed experimentally \cite{Walmsley03} and forms the basis of our characterization technique.
\bibitem{sign} We ascribe $+$ ($-$) index to the critical velocities for rotation in the same (opposite) direction as that of ${\bf \Omega}_0$ at which the texture was created.
\bibitem{erratum} This interpretation of the nature of $v_{\rm c}$ differs from that given in \cite{Walmsley03} in which on the basis of the proximity of $v_{\rm c+}= 0.6$\,mm\,s$^{-1}$ (for the 0.26\,mm slab) with the $v_{c0}\simeq 1$\,mm\,s$^{-1}$ due to the bulk dipole-unlocking instability when $\bl \parallel \pm\mathbf{v}$ \cite{Kopu00}.
\end{thebibliography}
\end{document}